\documentclass[10pt,journal,compsoc]{IEEEtran}

\IEEEoverridecommandlockouts
\usepackage{cite}
\usepackage{amsmath,amssymb,amsfonts}
\usepackage{algorithmic}
\usepackage{graphicx}
\usepackage{textcomp}
\usepackage{xcolor}

\usepackage{balance}
\usepackage{MnSymbol}
\usepackage{algorithmic}
\usepackage[linesnumbered,lined,commentsnumbered]{algorithm2e}%
\usepackage{makecell}

\usepackage{mathtools}
\usepackage{cuted}
\usepackage[hidelinks]{hyperref}

\usepackage{arydshln, booktabs}
\usepackage{cellspace}

\usepackage{fancyhdr}
\definecolor{darkblue}{RGB}{0,0,128}

\def\BibTeX{{\rm B\kern-.05em{\sc i\kern-.025em b}\kern-.08em
    T\kern-.1667em\lower.7ex\hbox{E}\kern-.125emX}}
\begin{document}

\title{Efficient deadlock avoidance for \\2D mesh NoCs that use OQ or VOQ routers}

\author{Philippos~Papaphilippou, Thiem Van Chu\IEEEcompsocitemizethanks{\IEEEcompsocthanksitem Philippos Papaphilippou is with the School of Computer Science \& Statistics, Trinity College Dublin, Ireland
(E-mail: papaphip@tcd.ie). %
}
\IEEEcompsocitemizethanks{\IEEEcompsocthanksitem Thiem Van Chu is with Tokyo Institute of Technology, Japan (E-mail: thiem@artic.iir.titech.ac.jp). %
}

}

\IEEEtitleabstractindextext{
\begin{abstract}

Network-on-chips (NoCs) are currently a widely used approach for achieving scalability of multi-cores to many-cores, as well as for interconnecting other vital system-on-chip (SoC) components. Each entity in 2D mesh-based NoCs has a router responsible for forwarding packets between the dimensions as well as the entity itself, and it is essentially a 5-port switch. With respect to the routing algorithm, there are important trade-offs between routing performance and the efficiency of overcoming potential deadlocks. Common deadlock avoidance techniques including the turn model usually involve restrictions of certain paths a packet can take at the cost of a higher probability for network congestion. In contrast, deadlock resolution techniques, as well as some avoidance schemes, provide more path flexibility at the expense of hardware complexity, such as by incorporating (or assuming) dedicated buffers.

This paper provides a deadlock avoidance algorithm for NoC routers based on output-queues (OQs) or virtual-output queues (VOQs), with a focus on their use on field-programmable gate-arrays (FPGAs). The proposed approach features fewer path restrictions than common techniques, and can be based on existing routing algorithms as a baseline, deadlock-free or not. This requires no modification to the queueing topology, and the required logic is minimal. %
Our algorithm approaches the performance of fully-adaptive algorithms, while maintaining deadlock freedom. %

\end{abstract}

}
\maketitle

\section{Introduction}

A network-on-chip (NoC) is an interesting and diverse approach for interconnecting a high number of computing entities. With the increase of the number of entities in today's processors and their heterogeneity, NoCs have an increasing presence in research. This includes field-programmable gate arrays (FPGA), where NoCs are applied to prototyping, CGRA implementation \cite{9912662} or simply for connecting systems of smaller logic components. %

One of the fundamental challenges in NoCs are deadlocks, and this is usually solved at the routing level or/and at the flow control level. This work focuses on the former. At the routing level, deadlocks are avoided by ensuring that the paths produced by the routing algorithm do not form any cycles. At the flow control level, deadlocks are avoided by preventing router buffers from being allocated to packets in a way such that a dependency cycle of packets is formed. Note that in practice, this distinction may be less clear, as it can relate more directly to implementation, such as to divide the router logic into pipeline stages \cite{dally2004principles}.

While the NoC research mostly focused on virtual channels (VCs) for buffering, in FPGAs and embedded systems, it is also common to use NoC designs with routers based on output-queued (OQ) switches \cite{kapre2006packet, huan2012fpga} or input-queued (IQ) switches with virtual output queues (VOQs) \cite{wang2010performance, ahmed2014graceful}. These have a queue for every input-output combination (see figure \ref{firouters}). As a result, each router in the nodes of such NoCs has a queue per pair combination of the 5 directions (\{E, S, W, N, C\} for east, south, west, north and centre respectively), assuming a 2D-mesh topology. %

\vspace{0.3em}
\textit{Limitations in state-of-the-art:} While there is a plethora of research on deadlock freedom in NoCs with routers based on virtual channels (VCs), NoCs that have a VOQ-like queue organisation still rely on simpler routing algorithms to achieve deadlock avoidance. This is because there is no theoretical background to increase path freedom for approaching full-adaptivity. While it is possible to adapt some methods from VCs, they are costly to apply, such as by introducing numerous additional %
queues for implementing/emulating escape channels. This limitation has an impact on routing performance, while the studied queue organisation remains popular among FPGA and embedded applications \cite{kapre2006packet, huan2012fpga, tpds23switches}. 

\vspace{0.5em}

\textit{Insights:} The main idea of this paper is that in NoC routers with VOQ-like queue organisation (one queue per input-output pair), deadlock avoidance can be achieved with more path flexibility than traditional models by \emph{calculating a worst case occupancy of specific queues}. This is based on the observation that \emph{this queue topology already implies turn information, which can be exploited to relax the turn model} for building more flexible deadlock-free routing algorithms. 

\vspace{0.5em}

\begin{figure}[h!]
\centering
\includegraphics[width=0.47\textwidth, trim=0 0 -3 0]{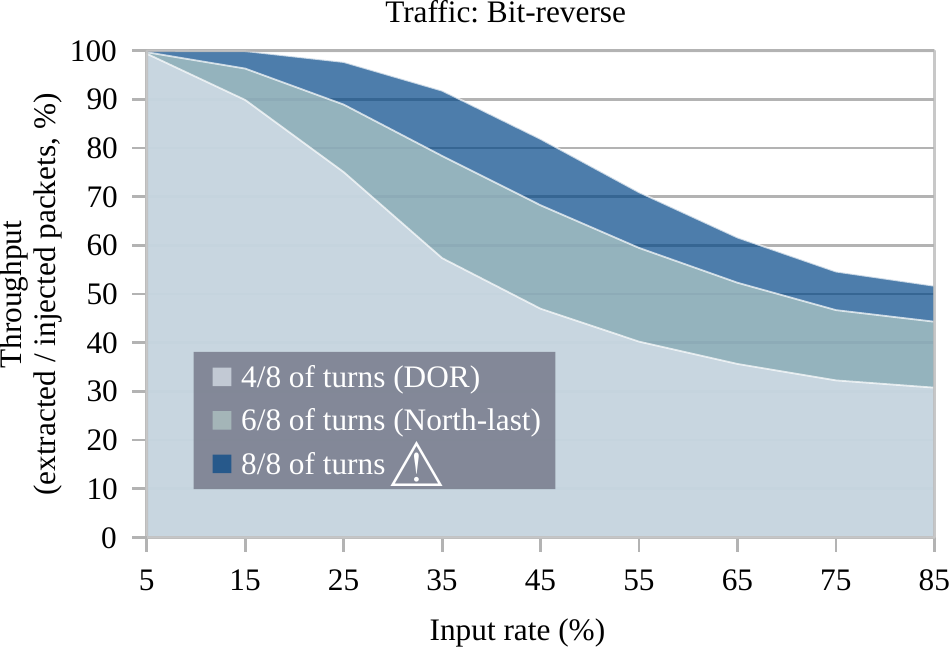}
\caption{NoC throughput under different quantities of turn restrictions}\label{fimot}
\end{figure}

\textit{Motivation:} In order to demonstrate the importance of path flexibility, figure \ref{fimot} presents simplified results from a \(8\times8\) NoC simulation with output-queues under 3 routing algorithms. The first algorithm is dimension order routing (DOR) which limits 4/8 of the possible turns, the second is ``north last" which forbids 2/8 of the available turns. The third one, however, does not have any turn limitation, but has a potential for creating deadlocks. The traffic model for this example was purposely selected to produce lower probability for deadlocks, in order to show the potential impact of path flexibility on performance. 

For instance, for a 55\% input rate of bit-reverse traffic \cite{dally2004principles}, DOR yields 
a throughput (extracted over injected packets) of 40\%, while the heuristic-based ``north-last" and the full-freedom ones yield 59\% and 72\% respectively. The goal is to approach the performance of the latter, but with deadlock freedom.
Note that these numerical values are case-specific. The throughput metric is simplified for the introduction here. %
See section \ref{routper} for more detailed discussion and simulations from which this example is based on.
\vspace{0.5em}

\textit{Contribution:} The proposed solution is a hybrid approach that solves the deadlock problem more efficiently than traditional routing algorithms. It gives the illusion of full freedom to any routing algorithm, deadlock-free or not, by using it by default, while intervening with a fallback algorithm when needed. %
A fallback algorithm such as XY or YX dimension order routing (DOR) is activated only locally and when deemed necessary according to the freedom condition. The freedom condition requires minimal information to decide if a packet can perform an originally restricted-turn. This is achieved without assuming additional queues (such as escape channels \cite{verbeek2011necessary}, interface buffers \cite{farrokhbakht2021pitstop}), deadlock detection \cite{ramrakhyani2018synchronized} and recovery routines \cite {farrokhbakht2021pitstop}, misrouting (non-minimal path), packet reordering and global knowledge that are usually found in the literature on NoCs with input-queued VC routers. An intentional but indirect outcome of this paper is also the increased routing performance of example hybrid algorithms exploiting the additional flexibility of the formally proven deadlock-avoidance technique. The evaluation framework is open source.\footnote{Source available: \url{https://philippos.info/deadlock}}

Following is background information (section \ref{backr}) on the related router architectures and the baseline deadlock avoidance model. Section \ref{sol} introduces the proposed algorithm, while section \ref{proof} is a proof for the correctness of the proposal. The evaluation (section \ref{eval}) includes simulations of example routing algorithms based on the proposed approach, as well as a study on the router resource utilisation and implementation efficiency. Finally, the paper concludes with related work (section \ref{rlw}), discussions on future work (section \ref{ftw}) and a conclusion (section \ref{conc}).

\section{Background}\label{backr}

\subsection{Network-on-chip routers}

NoC routers are essentially switches, forwarding packets from port to port. The most common NoC routers are based on an input-queued switch with virtual channels (VCs). On each port (\{E, S, W, N, C\} for east, south, west, north and centre), the input is connected to a demultiplexer splitting the packets/flits into a fixed number of VCs (buffers). A virtual channel allocation scheme is one of the first decisions a NoC router makes on packet arrival, which can also relate to quality of service (QoS). 

\begin{figure*}[h!]
\centering
\includegraphics[width=0.75\textwidth, trim=0 10 0 0]{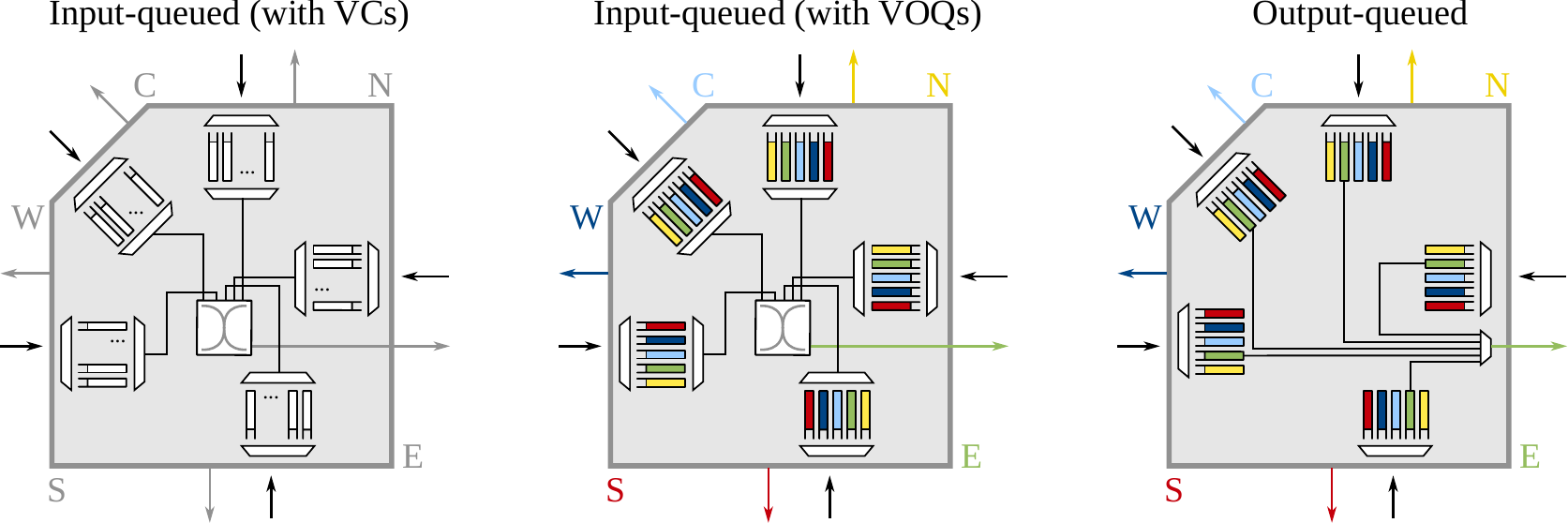}
\caption{Input and output-queued NoC routers. Colour-coded queues feed the output port of the corresponding colour. The non-coloured queues rely on virtual channel (VC) allocation, hence each of their packets can be destined to different output ports. The outputs to ports other than E are abstracted for simplicity.}\label{firouters}
\end{figure*}

Each group of VCs are then demultiplexed into the assumed crossbar (functionally a superset of permutation networks) connected to the outputs (ports), resulting in a \(5\times5\) switch. The buffering is necessary to facilitate the temporary storage of incoming packets until a passable matching is achieved between the output ports and all virtual channels. A matching is calculated on every fixed period of time and is used by the crossbar. See figure \ref{firouters} (top left) for an illustration of a router with a number of VCs near each of its inputs. %

\subsubsection{Input-queued routers with VOQs} 

Instead of virtual channels (VC), virtual output queues (VOQs) can be used to implement switches, and are especially common in network switches. There is a queue for every input-output port combination. Thus, there are \(P\) groups of \(P\) VOQs, resulting in \(P^2\) total buffers for a \(P\)-port switch (in 2D-mesh NoCs, \(P=5\)).

A disadvantage of virtual output queues over virtual channels in NoCs is that they work as a static virtual allocation scheme in a switch with 4 or 5 virtual channels. This essentially means that the queue utilisation may be less efficient, and the buffer space is generally higher. It is also more challenging to reduce the number of those queues (one way is to restrict routability, see the discussion of section \ref{resul}). Additionally, supporting quality of service (QoS) can challenge scalability, due to the increased queuing requirements. 

However, as both VCs and VOQs support memory sharing per queue group, their differences can become less significant according to the implementation, also considering that 4 VCs may already be a norm in modern processors \cite{dai2022full, ma2012whole}. %

\subsubsection{Output-queued (OQ) routers} 

Another approach to NoC router implementation is output-queues. They eliminate the use of a crossbar in favour of simpler logic. In this case, the queue organisation is the same with virtual output queues, and is illustrated in figure \ref{firouters} (bottom). 

On FPGAs, this is a prominent NoC router architecture \cite{fpt21switch}, as with the split-merge switch \cite{huan2012fpga,kapre2006packet}. The split-merge switch is an output-queued switch meaning that, upon arrival, the incoming packets are immediately split across queues according to the destination port. Those queues are then grouped based on their destination port, and are multiplexed only once. This results in low-complexity and/or highly-pipelinable logic, the split and merge units, which are here equivalent to demultiplexers and multiplexers respectively. 

There is no need for an expensive scheduling algorithm, such as by featuring an iterative approach to perform well. There is only arbitration near the output, for every output port, which can be fulfilled by only using priority encoders. This also results in high scheduling performance, as it naturally provides more connectivity combinations than input queued for the same queue topology. This is because of the absence of additional arbitration steps near the inputs, which would otherwise serialise dequeuing from the output queues coming from the same input port.

A potential disadvantage of output-queued switches as NoCs routers is the limited scope for memory sharing, the associated cost of which also relates to QoS support. Both  memory sharing and QoS are seemingly less popular on FPGAs, as simpler designs are preferred \cite{huan2012fpga,tpds23switches}.

Our proposal provides the theoretical foundations to provide more flexibility in routing for such VOQ-like queue topologies, instead of relying on worse-performing traditional routing.

\subsection{Turn Model} %
\label{sec:turn-model}

\begin{figure}[h!]
\centering
\includegraphics[width=0.39\textwidth, trim=0 0 0 0]{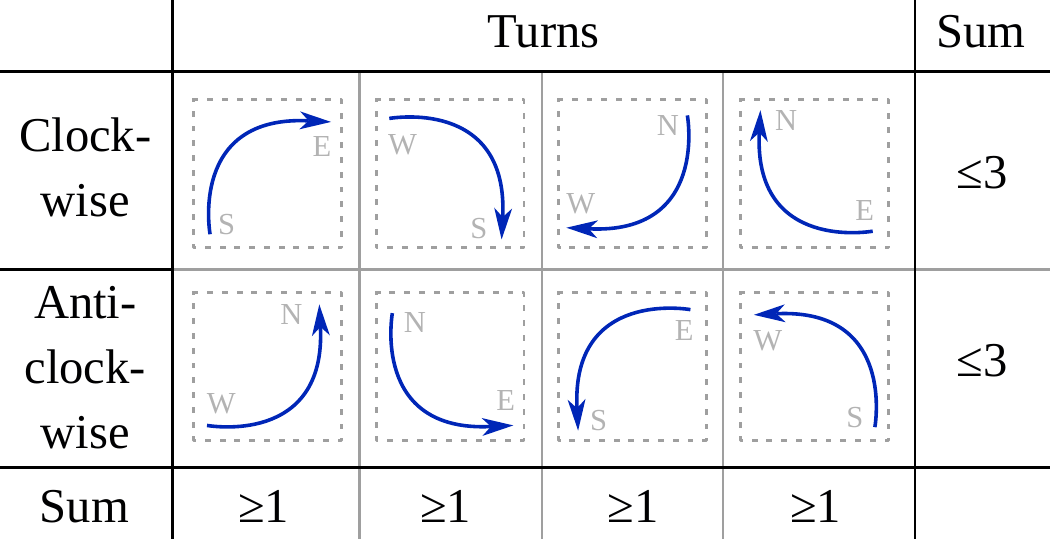}
\caption{Turn model for deadlock avoidance}\label{fiturn}
\end{figure}

The turn model can be used to create routing algorithms based on turn restriction to avoid forming any possible cyclic dependency \cite{glass1992turn}. The resulting set of possible algorithms can be summarised by the rules illustrated in figure \ref{fiturn}. There are clockwise and counterclockwise turns for which at least one turn from both must be forbidden. However, for each diagonal direction there must be at least one way for packets to travel, hence the ``\(\geq1\)'' sum rule per column in the figure. This model is a superset of the older ``dimension ordered routing'' (DOR), i.e. XY DOR for first traversing along the x-axis (no %
``\(\rcurvearrownw,\lcurvearrowne,\lcurvearrowsw,\rcurvearrowse\)'' turns) and YX DOR (no %
``\(\lcurvearrownw,\rcurvearrowsw,\rcurvearrowne,\lcurvearrowse\)'' turns).

When reducing the restrictions as much as allowed by this model, there needs to be one clockwise and one counterclockwise turn restriction, out of which they are not on a common path, as explained above. This results in 12 possible combinations for the forbidden turns (permutations of 2 from 4 (i.e. $P_{(4,2)}$), as repetition from the table columns would lead to forbidden destinations). In other words, the turn model gives 12 turn-restriction-based algorithms. However, if a rotated mesh is considered equivalent, this reduces the number of algorithms to only 3: %
``west-first'' (no %
``\(\rcurvearrownw,\lcurvearrowsw\)'' turns), ``north-last'' (no %
``\(\rcurvearrownw,\lcurvearrowne\)'' turns) and ``negative-first'' (no  %
``\(\rcurvearrownw,\lcurvearrowse\)'' turns) routing algorithms.

\subsection{Deadlocks}

The deadlocks that happen in NoC routers with a VOQ or OQ queueing topology are a bit less trivial than the classical example paths on a simple \(2\times2\) mesh \cite{lopez1998very}. In such a deadlocked \(2\times2\) mesh with a single queue (of length 1) per link per node, the remaining packets all have 2-hop path, as 1-hop paths would have been consumed directly by the destination, and 3-hop paths would have been U-turns (this paper only studies minimal path routing). In this case, there is a cyclic dependency on the four buffers. In the OQ and VOQ case, however, it is impossible to create a deadlock on a \(2\times2\) mesh, as all of the four different 2-hop paths never pass from the same queue. %

\begin{figure}[h!]
\centering
\includegraphics[width=0.35\textwidth, trim=0 0 0 0]{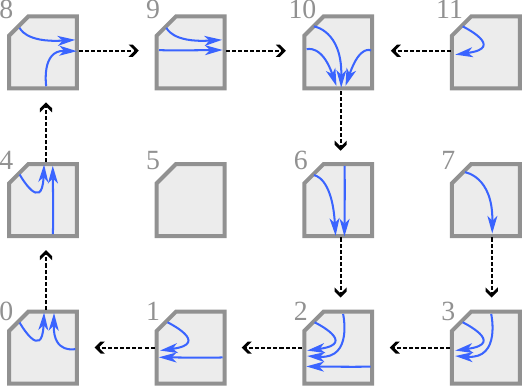}%
\caption{Extended deadlock state example from a simulation of a \(3\times4\) NoC with OQs. %
}\label{fivd}
\end{figure}

Figure \ref{fivd} illustrates an extended deadlock state (i.e. also including blocked queues not necessarily from the cyclic dependency) %
that is observed in a simulation of \(3\times4\) mesh NoC with OQs. The OQs are of length 2 and the simulation uses a routing algorithm without deadlock avoidance. The numbered nodes represent the NoC nodes, while the blue arrows %
represent the filled output queues. %
The start and target positions of the blue arrows denote the source and destination of single turns (e.g. ``$\lcurvearrowne$" for the SE turn), as (V)OQs are associated here with turns.  
The dashed arrows represent queue dependencies on filled queues. They do not necessarily show which queues are involved (the heads of two queues can target the same destination queue), though here it is more apparent, given that this simulation is in a saturated state.

\section{Proposed algorithm}\label{sol}

The proposed algorithm is bimodal, and avoids deadlocks by selecting between a base algorithm and a fallback algorithm based on the freedom condition \(F\) (section \ref{frco}) for every individual next hop decision. The fallback routing algorithm in our case shall follow the turn model, and can be for example XY DOR, ``west-first'', etc. The algorithm \ref{alg2} presents the distributed algorithm among the NoC's nodes that relates to the routing decisions of OQ or VOQ-based NoC routers.

\begin{algorithm}[h!]
\small
\LinesNumbered
\SetSideCommentRight
\SetKwComment{Comment}{$\triangleright$\ }{}
in \(inp=from(p)\)\Comment*[r]{\hspace{-1em} the input port receiving \hspace{-1em} \(p\)}
out \(sel\)\Comment*[r]{the output port selection}
 \While{forever}{
 receive (\(p\));\\
\eIf {F(p)==True}{
        \(sel\) $\leftarrow$ \(base\_algorithm(p)\);\\        
}{
        \(sel\) $\leftarrow$ \(fallback\_algorithm(p)\)\Comment*[r]{Turn model}
}    
OQs[\(inp\)][\(sel\)].\(enqueue(p)\);
 }
\caption{Proposed %
routing algorithm for (V)OQs}\label{alg2}
\end{algorithm}

As the goal of this algorithm is to provide path flexibility, such as for better routing performance, the base algorithm is expected to allow a superset of the turns allowed by the fallback algorithm. The base algorithm can be any arbitrary routing algorithm, deadlock-free or not, which is also very useful for adopting algorithms that would otherwise require a specialised queue organisation for achieving deadlock freedom. Some examples include O1-Turn and LEF \cite{lef} that assert virtual channel allocation requirements. Like with turn model algorithms, the flexibility of the base algorithm is also appropriate for adaptive routing, where %
the routing decisions are based on heuristics such as for estimating network congestion. %

Sections \ref{frco}  and \ref{fradapt} provide a detailed description of the proposed freedom condition. %

\subsection{Assumptions}\label{assum}  %

The proposed approach is studied under the assumptions %
listed in table \ref{tab0}. %
The table classifies these assumptions based on whether they are a requirement of the proposed algorithm (``fundamental''), or if they are design choices serving simplicity or practicality (``adaptable''). The discussion of section \ref{ftw} elaborates on some trivial cases for generalising some of the ``adaptable'' attributes.

\begin{table}[h!] 
\caption{Assumptions} 
\label{tab0} 
\centering
  
\setlength{\tabcolsep}{4.5pt}
\begin{tabular} {r | c c }
Attribute&Value&Comments\\
\multicolumn{1}{l;{1pt/2pt}}{}&&\\
\multicolumn{1}{l;{1pt/2pt}}{\textit{Fundamental:}}&&\\
NoC topology&2D mesh&\\
Queue organisation& \thead{OQ or\\ VOQ}&\thead{up to \(5\times5\) virtual/physical \\queues from (N, W, S, E and C)\(^1\)}\\
Variable-size packets&supported&\thead{head flit carries the packet size\\(elaborated in sections \ref{frco} and \ref{ftw})}\\
\multicolumn{1}{l;{1pt/2pt}}{}&&\\
\multicolumn{1}{l;{1pt/2pt}}{\textit{Adaptable:}}&&\\
Multi-flit packets&supported&\\
Allowable paths&\thead{minimal\\(shortest)}&\thead{Manhattan distance \\(i.e. \(\lvert x_1-x_2 \rvert + \lvert y_1-y_2\rvert\))} \\
Misrouting&no&no temporary redirections\\
Interface queues&no&\\
Flow control&\thead{ready signal\\ per port %
}&\thead{5-bit based on queue occupancy \\ per (N, W, S, E and C) turn\(^{1}\)}\\%
Re-allocation scheme&WPF \cite{ma2012whole}&\thead{non-empty channels can receive\\ new packets, if a tail flit arrives} \\
Processing latency&1&router implementation-specific\\
\multicolumn{3}{c}{}\\
\multicolumn{3}{c}{\(^1\){fewer in practice, as minimal path contains no U-turns etc.}}\\
\end{tabular}

\end{table}

The series of events when a node receives a packet is the following: routing (proposed algorithm), virtual channel allocation (fixed because of the (V)OQ organisation%
), scheduling (including for the whole crossbar when using VOQs \cite{tpds23switches}) and finally switch traversal. The routing decisions of an upstream node are assumed not to influence the routing decisions of a downstream node. %

The proposed approach supports multi-flit packets. The paper focuses mostly on single or few-flit packets, a common case for NoCs in processors \cite{ma2013novel}. This is also the case with NoCs working as system interconnects inside FPGAs. Hence the use of whole packet forwarding (WPF) that avoids unnecessary blocking of the buffers (see table \ref{tab0}). A simple scheduling assumption for the output arbiters is required to give priority to flits of the same packets (e.g. before applying round-robin) to avoid packet overlapping. %

In the nomenclature,
the term ``turn'' is also sometimes used for forwardings from and to a port that is on the opposite side (going straight), as well as to the node's centre (``C''). These are considered legal %
in accordance with
the turn model. %
The centre in this case is a ``sink'', being able to consume packets directly \cite{holsmark2009deadlock, duato1993new}. A node's centre as a consumer cannot contribute to a deadlock, as every inserted packet is eventually consumed in a dedicated (virtual) output queue. %
Also, the turn model does not consider straight forwardings for deadlocks, though different models can also break circles in straight sections of a packet's path. %

\subsection{Freedom condition}\label{frco}
The freedom condition (\(F\)) is a sufficient %
condition for the avoidance of deadlocks, and is applied on packet arrival into any non-sink node (sink is the final destination). %

The main idea of \(F\) goes as follows. If an incoming packet \(p\) can do a clockwise or a counterclockwise turn on the next hop, and those turns are considered restricted based on the fallback algorithm (e.g. ``north-last"), we check the worst case occupancy for that queue of the next hop router. This is achieved by summing up the contents of all queues that feed into that queue (only the packets that can go into that queue), plus its own contents, and checking whether the addition of the incoming packet would cause an overflow in the worst case. The worst case is for all packets in the sum to end up in the restricted-turn queue and that for any reason it stops being consumed in the meantime. The time frame for the worst case occupancy is for until packet \(p\) manages to be enqueued into the restricted turn/queue. Whenever \(F(p)=\textrm{\textit{False}}\), \(p\) takes an alternative queue/output port/direction, which ensures that \(p\) will not be able to make a turn not following the turn model of the fallback algorithm in the routing step of the next hop.

\begin{figure}[h!]
\centering
\includegraphics[width=0.43\textwidth, trim=0 5 0 0]{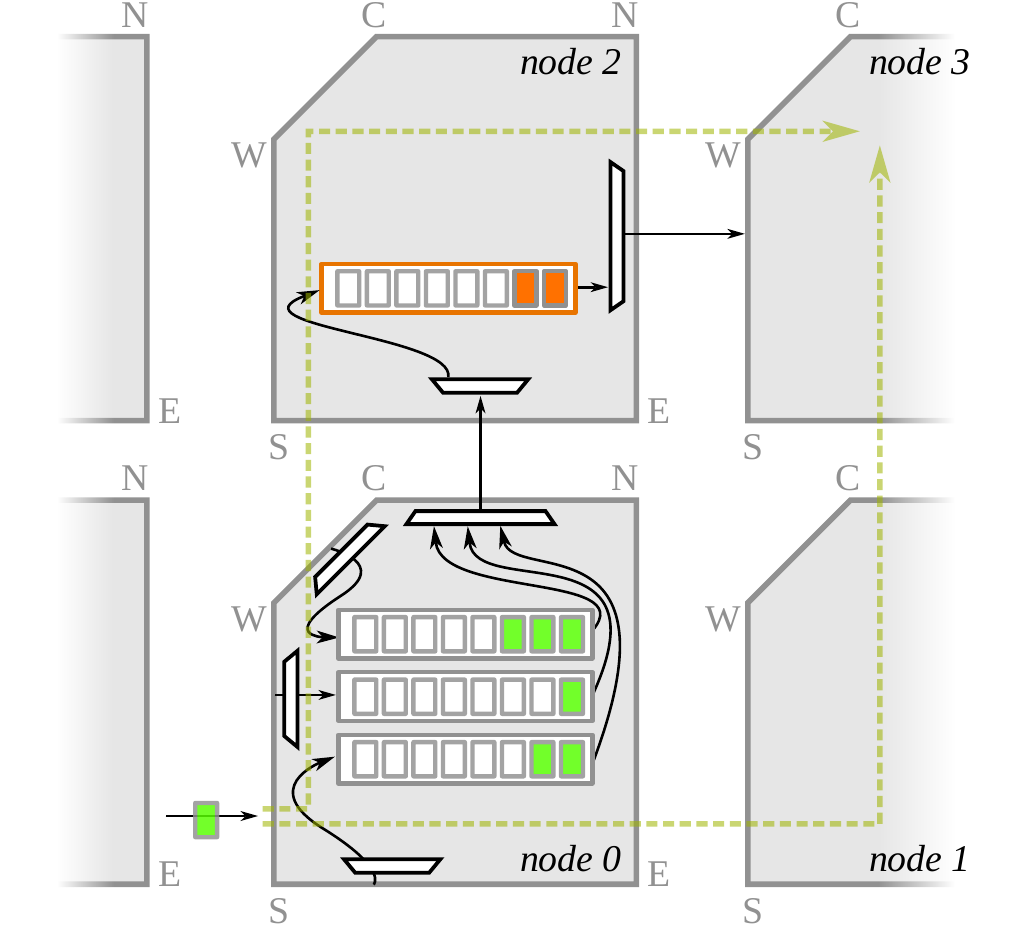}
\caption{Example NoC state with single-flit packets and OQ (of depth 8)}\label{fiex}
\end{figure}

Figure \ref{fiex} illustrates the main idea with an example NoC state that can benefit (avoid a deadlock) from the use of the freedom condition. %
An incoming packet to node 0 has node 3 as a destination. This packet can be routed to two possible queues, one for going through node 2 (middle queue in node 0) and one for going through node 1 (not shown). The depicted queues in node 0 with green packets are all the queues that feed node 2 from its south port. Given that the fallback routing algorithm is based on the turn model, and the originally-forbidden turn is on the path through 2 (``\(\lcurvearrowne\)" queue/turn in node 2, with red hint), we notice that the packet may not be able to be accommodated, if the forbidden path is followed. Assuming all packets (green) in those queues target the red queue, i.e. based on packet destination, the worst case occupancy for the red queue is 8 (2+1+3 from node 0 and 2 from node 2). This will happen, for example, under the current queue states, if the red queue stops being consumed in the meantime. Thus, the incoming packet could cause a stall, if it follows the path through 2, and it would be safer to follow the path through node 1. The receiving queue in node 1 (``\(\rcurvearrowne\)" turn) is not forbidden by the turn model, as it is the counterclockwise direction-equivalent of the forbidden one ($\geq1$ requirement in section \ref{sec:turn-model}), and therefore will not cause a cyclic dependency on its own. %

Let \(p\) be a packet or flit received from any port \(from(p) \in \{E, S, W, N, C\}\). %
Let \(sel \in \{E, S, W, N, C\}\) be the selected direction of the proposed routing algorithm.  based on the result of the freedom condition. %

The freedom condition is only useful whenever the next hop implied in \(sel\) could make \(p\) land in a queue of the next hop that would be a restricted turn based on the turn model. \(next(p)\) is the set containing all possible queue destinations of \(p\) inside the next-hop's router, and in practice here it also follows no-misrouting. %
When \(p\) arrives at the next hop, according to its designated final destination it can go straight (``\(\nearrow\)''), to the centre, clockwise (``\(\lcirclearrowright\)'') or counterclockwise (``\(\rcirclearrowleft\)''). Based on our assumptions the latter two are the only ones that can be forbidden turns, and are mutually exclusive. %

The piecewise function of the freedom condition is as follows: %
{\small
\(F(p) \equiv \)  
\[\begin{cases} 
      size(p) +  %
      occp(q'_\lcirclearrowright)
      +\sum_{d \in \{C, \nearrow, \rcirclearrowleft\}} 
      occp(q_d)
      \leq cap(q'_\lcirclearrowright), \ \ \ \ \ \ \ \ \ \ \ \ \ \ \ \\[6pt]
      \hfill  q'_\lcirclearrowright \in next( p) \land  \neg  turn(q'_\lcirclearrowright)\\
       \\
      size(p)
	+occp(q'_\rcirclearrowleft)
      +\sum_{d \in \{C, \nearrow, \lcirclearrowright\} }  
      occ'(q_d)
      \leq cap(q'_\rcirclearrowleft), \\[6pt]
      \hfill q'_\rcirclearrowleft \in next(p) \land  \neg  turn(q'_\rcirclearrowleft)  \\
      \\
       \textrm{\textit{True}}, 
      \hfill Otherwise\\
\end{cases}
\]
}

where %
\(cap (q)\) is the %
capacity of queue \(q\), %
\(q_d\) is a queue of the current node (receiving \(p\)) that receives packets %
from direction \(d\), while \(q'_d\) is a queue of the next hop as pointed by \(sel\).
\(turn(q)\) indicates that the queue \(q\) corresponds to a legal turn based on an algorithm derived by the original turn model. 

The occupancy of queue \(q\) is calculated as 
\(occp(q)=\sum_{p' \in q}{size(p')}\), which counts multi-flit packets rather than the physical occupancy. %
The \(size(p)\) function indicates the number of flits a packet \(p\) consists of. This is useful in the cases where multi-flit packets are allowed, where this function is only useful on the head flits of packets. The availability of this information is assumed. %
For multi-flit packets the head flit shall mention the number of flits for the whole packet. For non-head flits of a packet \(p\), \(size(p)=0\), since the final packet occupancy for multi-flit packets should be known by the time their first flit is accommodated. %

\subsection{Freedom condition adaptation}\label{fradapt}

In order to simplify the presentation of the freedom condition also based on implementation practicality, this subsection provides an adaptation example. The section provides a simplified sufficient condition \(F'\) based on system implementation assumptions and a higher degree of approximation for the queue occupancy, but with the same worst case (i.e. \(F'(p)\rightarrow F(p)\)).

The size of the packets (\(size(x)\) function) is replaced by 1, as only single-flits are considered for brevity. A related modification is to replace the packet/flit counts having a potential queue target inside the next hop with the entire (physical) occupancy of each of the corresponding queues (denoted by \(occ(x)\) for each queue \(x\)). Although there can be an overhead in the path flexibility the algorithm provides, it could also be implemented more efficiently as the queue length circuitry is likely to be already existent. %

Another adaptation is that the ``north last'' routing algorithm is selected as the fallback condition (or a restriction subset that still follows a turn model). By having the same output port (``N'') as the potential direction for the start of both forbidden turns for the next hop (SW, SE, i.e. ``\(\rcurvearrownw,\lcurvearrowne\)''), there needs to be less logic for the queue capacity checks in total. This is because \(q_C, q_\nearrow, q_\lcirclearrowright\) and \(q_\rcirclearrowleft\) represent the same queues for the turns SW and SE, happening at \(q'_{\rcirclearrowleft}\) and \(q'_\lcirclearrowright\)  respectively in the next hop across the N direction. Under this assumption, the counterclockwise, straight and clockwise of the current node in \(F\) become the WN, SN and EN queues (``\(\rcurvearrowne, \uparrow, \lcurvearrownw\)'') and for the next hop the SW, SN, SE (``\(\rcurvearrownw, \uparrow, \lcurvearrowne\)'') respectively.

The link overhead in this case is two southerly wires between each consecutive neighbouring nodes in the \(y\)-axis, of bit widths equal to \(\lceil \log_2(cap(q'_\lcurvearrowne))\rceil\) and \(\lceil \log_2(cap(q'_\rcurvearrownw))\rceil\) correspondingly. This metric makes the simplifying assumption that the occupancy of the next hop can be provided within the same cycle. A practical implementation with a credit system, such as with pipelined router implementations, is also likely to use fewer wires between the nodes.

\(F'(p) \equiv        \)
\[ \begin{cases} 
      1 +  occ(q'_\lcurvearrowne) +\sum\limits_{d \in \{C, \uparrow, \rcurvearrowne\} } occ (q_d)  \leq cap(q'_\lcurvearrowne), &  q'_{\lcurvearrowne} \in next(p) \\
       \\
      1 +  occ(q'_\rcurvearrownw) +\sum\limits_{d \in \{C, \uparrow, \lcurvearrownw\} } occ (q_d)  \leq cap(q'_\rcurvearrownw), &  q'_{\rcurvearrownw} \in next(p) \\
\\
       \textrm{\textit{True}}, & Otherwise\\
   \end{cases}
\]

\begin{figure}[h!]
\centering
\includegraphics[width=0.45\textwidth, trim=0 7 0 0]{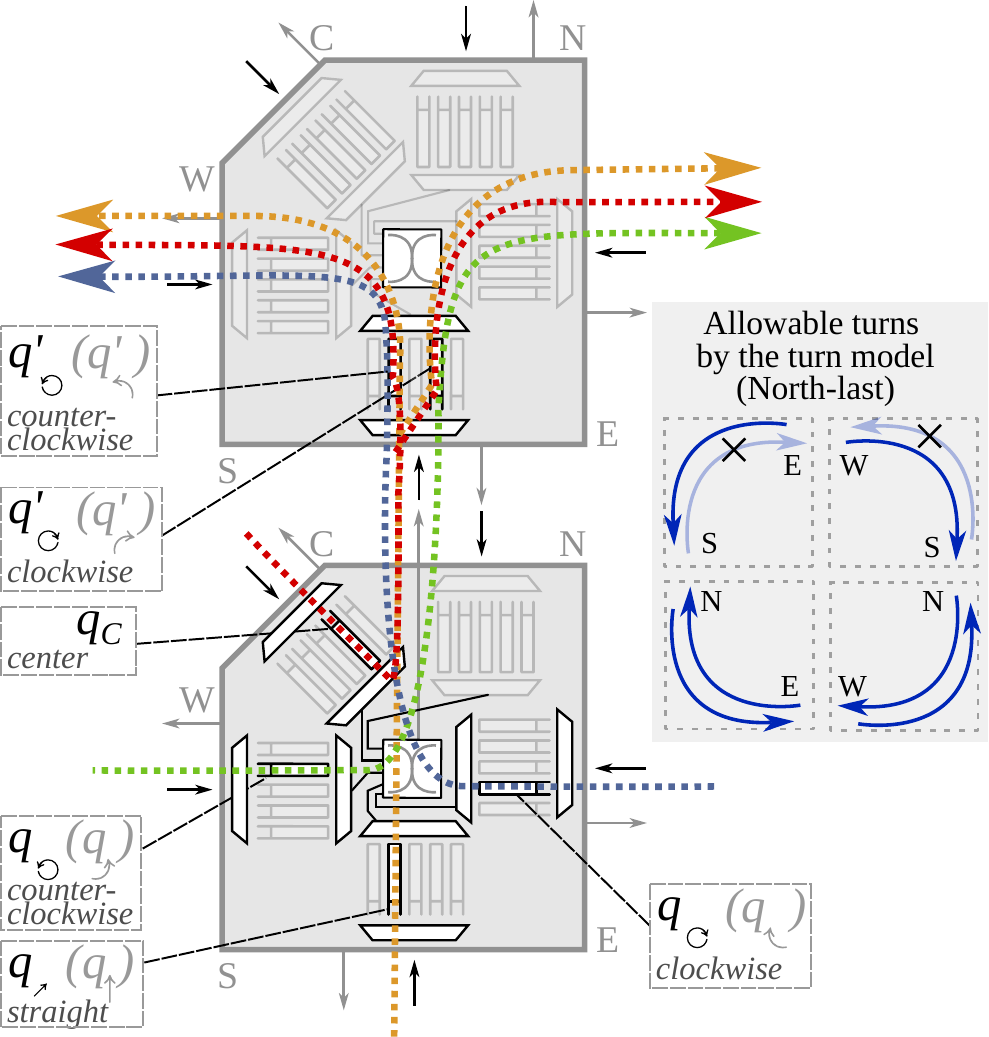}
\caption{Packet paths involving originally-forbidden turns on the next hop}\label{fiproof}
\end{figure}

Figure \ref{fiproof} illustrates the potential paths that involve non-allowable turns/queues by ``north-last'' as the fallback routing algorithm, which restricts SW and SE turns. When a packet can make such turns in the next hop (associated with the queues \(q'_{\rcurvearrownw}\) and \(q'_{\lcurvearrowne}\)), then exactly one of the first two pieces of the piecewise function \(F'\) is used.

For the sake of notation simplicity both \(F\) and its variation \(F'\) are only valid when applied ``serially'' on the set of incoming packets per cycle (but still being a combinational circuit operating in the same cycle). In practice, their computation can be implemented in parallel for each input port, but %
there needs to be a simple arbitration step for synchronisation, such as with a priority encoder among the input ports. This is for whenever two packets can compete for the same potential queue of the next hop (%
packets from different ports/directions are still placed in different queues).

\subsection{Novel routing algorithms}\label{novela}

Two example routing algorithms are provided for the purposes of evaluation, and are based on the proposed bimodal algorithm. The first one is  ``XY/Adaptive" and uses XY DOR as a fallback algorithm. This fallback algorithm is equivalent to the restrictions as found in 
``north-last" in our arrangement (adaptation of section \ref{fradapt}), as \(F'=\textrm{\textit{True}}\) for south ports, since they cannot form a forbidden turn as per our base turn model. The ``Adaptive" component of ``XY/Adaptive" is full-freedom, i.e. there is no turn restriction and the routing decisions follow a heuristic.

As adaptive routing algorithms like ``north-last" introduce a degree of turn freedom, heuristics are used to avoid congestion. Throughout the entirety of this study, the adaptiveness heuristic is consistent for all experiments. It is enabled where applicable, including for the novel ``XY/Adaptive". The higher priority is given to the turn for which the direction maps to the queue with the least occupancy. This is conventional, but in this way, the information used by the heuristic is also local to the node. The occupancy of each local queue is still indirectly associated with the congestion in the corresponding next hops.

The second proposed algorithm based on the freedom condition \(F'\) is ``XY/O1-Turn", which alternates between XY DOR and O1-Turn \cite{o1turn} per packet arrival.  O1-Turn randomly selects XY or YX DOR for the whole path of a packet, and does not provide adaptiveness. O1-Turn is originally designed for routers with virtual channels and achieves deadlock freedom by using separate queues/channels for XY and YX packets \cite{lef}. Therefore, this is an example where the proposed technique is used to adapt an algorithm to a VOQ-like queue organisation, without the need to double the storage requirements. In this case, the aforementioned heuristic is never consulted, as XY DOR acts as a fallback algorithm when O1-Turn's decision involves the north port and \(F'\) considers it unsafe.

\section{Proof}\label{proof}

In order to prove that the algorithm in \(F\) is always deadlock-free, a proof by %
contradiction is provided. %

Let \(c\) be a cyclic dependency \cite{duato1995necessary} between \(n\) buffers that has been allowed by the routing algorithm at time \(t\).
\begin{align}
c=\{q_0, q_1, ..., q_{n-1}, q_0=q_n\}
\end{align}
 As this is a deadlock, all queues in the cycle are full, with each head packet only able to be served by queues of the subsequent node (including the subsequent queue in the cycle). That is 
 \begin{align}
 occ(q_i)=cap(q_i) \forall i \in \{0, 1, ..., n-1\},
\end{align}
where \(occ (q_i)\) and  \(cap (q_i)\) is the occupancy and capacity of queue \(q_i\) respectively. Additionally, 
 \begin{align}
q_{j+1} \in next(head(q_j))
  \end{align}
  for every \(j\) rotation %
  of \(i\), where \(next(head(q_j))\) is the set %
containing all possible buffer destinations of the head of queue \(q_j\). Note that as long as a packet is in a queue, the direction for the next hop (corresponding to %
a specific node) is already determined, but the queue placement in the next-hop node will still be decided upon arrival %
at time \(t+1\). %

For every \((q_j, q_{j+1})\) pair, a head packet proceeds from \(q_j\) to \(q_{j+1}\) if and only if the algorithm considers the packet making a legal forwarding to \(node(q_{j+1})\) (for deadlock avoidance), and the \(node(q_{j+1})\) notifies that it will be able to accommodate it on cycle \(t+1\). %

As each queue represents a turn (a permutation of 2-selection from \{E, S, W, N, C\}), based on the turn model, it is impossible for the cyclic dependency to be based only out of turn/queues being inline with %
the turn model. That is \(\neg( \forall i, q_i \in c \land   turn(q_i))\), where \(turn(q_i)\) denotes that \(q_i\) follows the turn model. In other words, there is at least one queue/turn not following the turn model. That is, for \(c\) to be able to be formed,
\begin{align}
\exists i, \ q_i \in c \land  \neg turn(q_i). %
\end{align}

Similarly, based on the turn model, as summarised in figure \ref{fiturn}, the forbidden turns alone are also not able to form the circle \(c\) either, as they will consist of a strict subset of the 4 turns in the clockwise direction and a strict subset of the 4 turns in the counterclockwise direction. Therefore, 
\begin{align}
 \neg(\forall i, q_i \in c \land   \neg turn(q_i)) \\
 \leftrightarrow        \exists i, \ q_i \in c \land  turn(q_i).
\end{align}

From (4) and (6), there is at least one consecutive queue pair that consist of one queue following the turn model and one that does not, i.e. \(\exists l=(i+k)\ \mathrm{mod}\ n, k \in \mathbb{Z},\)
\begin{align} turn(q_l) \land \neg turn(q_{l+1}) \end{align}

As this pair also denotes a dependency ( \(q_{l+1} \in next(head(q_l))\) ), at the time of insertion of the now \(head(q_l)\) to \(q_l\), the \(F\) condition has been met, i.e. %
\begin{align}
size(p)+ occp(q_{l+1}) +\sum_{d \in \{C, l, l'\}}{occp(q_{d})}  &\leq cap(q_{l+1}),
\end{align}
where \(q_{l'}\) is the other queue in the \(node(q_l)\) than \(q_l\) that can feed \(q_{l+1}\). One of them  (\(q_{l}\) and \(q_{l'}\)) is for the straight movement (\(q_\nearrow\), or \(q_\uparrow\) for \(F'\)). %
The other, based on \(F\), if \(q_{l+1}\) is clockwise (\(q'_\lcirclearrowright\), or \(q'_\lcurvearrowne\) for \(F'\)), it is counterclockwise (\(q_\rcirclearrowleft\), or \(q_\rcurvearrowne\) for \(F'\)), and vice-versa (if \(q_{l+1}=q'_\rcirclearrowleft\) (or \(q'_{\rcurvearrownw}\) for \(F'\)), then it is \(q_\lcirclearrowright\) (or \(q_\lcurvearrownw\) for \(F'\))). There is also \(q_{C}\) that starts from the centre of \(node(q_l)\). These queues (\(q_l\), \(q_{l'}\) and \(q_{C}\)) are the only queues that can feed \(q_{l+1}\) based on the assumptions, such as with no misrouting. %

First, in section \ref{sfpr}, the proof is done under the assumption that there is only a single flit per packet. Then, section \ref{mfpr} elaborates on the multiple-flit case.

\subsection{Single-flit packets}\label{sfpr}

 \begin{figure*}[h!]
\centering
\includegraphics[width=1\textwidth, trim=0 0 0 0]{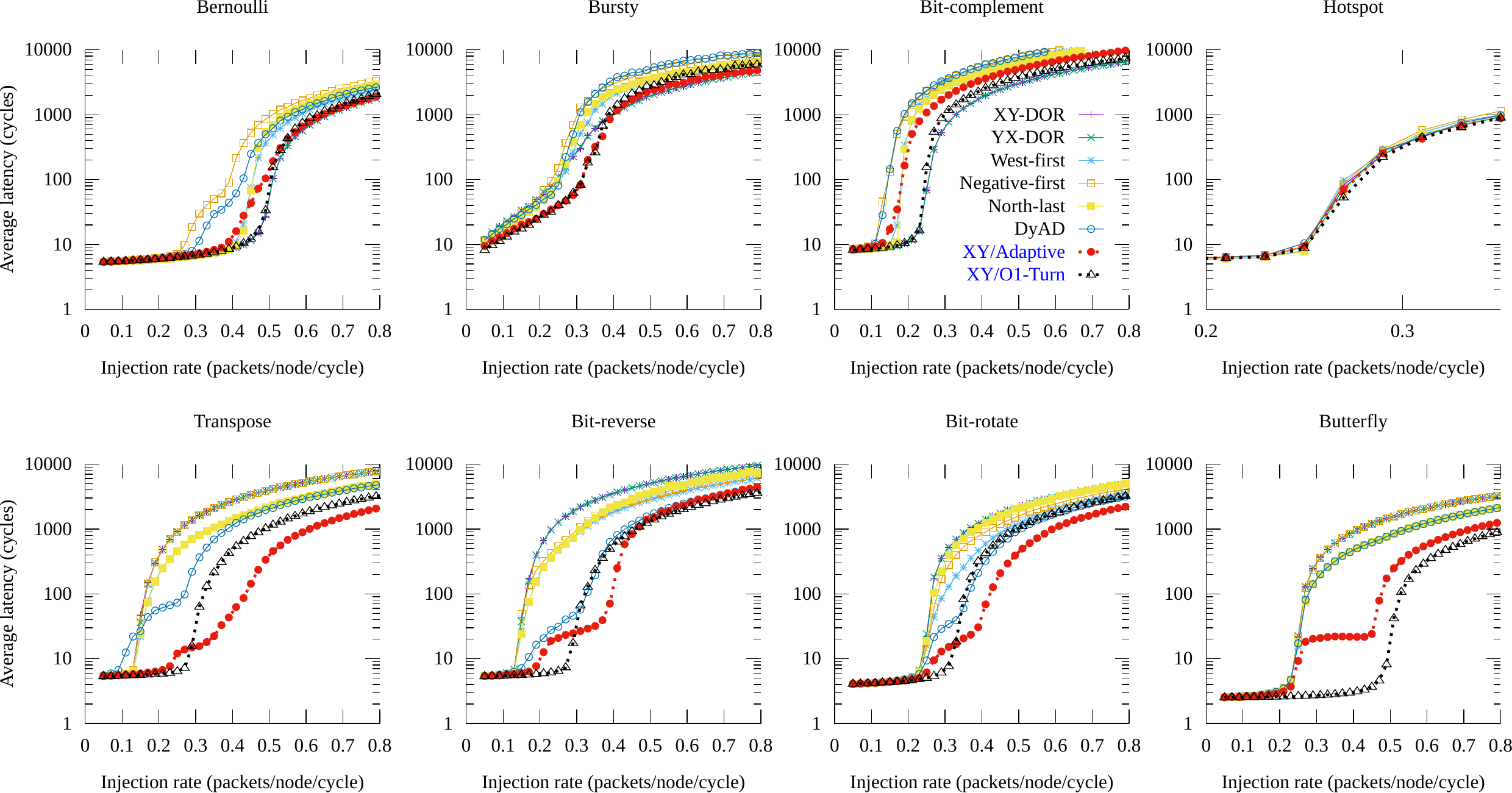}
\caption{Comparison of average packet latency using different routing algorithms under different traffic patterns. %
}\label{figlat}
\end{figure*}

As \(F\) is always followed, at the time \(head(q_l)\) was enqueued into \(q_l\), the value of \(F\) was \textit{True}, i.e.%
\begin{align}
1+ occ(q_{l+1}) +\sum_{d \in \{C, l, l'\}}{occ(q_{d})}  &\leq cap(q_{l+1}).
\end{align}

Following the worst case, %
it is assumed that
the queue \(q_{l+1}\) stops being consumed in the meantime for any reason, such as congestion. %
Any packet \(p\) that arrived to \(node(q_l)\)  after (the now) \(head(q_l)\) would have been one of the following cases based on the turn options allowed by the minimal path:\\[-6pt] %
\paragraph*{\hspace{-1em}\textit{A) \(p\) can land in \(q_{l+1}\)}}%
this is the case when \(p\) avoids the ``forbidden'' turn (\(q_{l+1}\)), since the freedom condition causes it to be routed it to its alternative direction.
As \(F(p)=\textrm{\textit{False}}\) based on the current occupancies, \(p\) will follow the fallback algorithm which will place it in a queue other than \(q_l\), \(q_C\) and \(q_l'\), as it will target a different node to \(node(q_{l+1})\). %
Since the turn model always provides full-routability without relying on non-allowable turns, %
\(p\) will simply follow the fallback routing algorithm, which will ensure a next hop without a queue violating the turn model. %

\paragraph*{\hspace{-1em}\textit{B) \(p\) cannot land in \(q_{l+1}\), but still passes through \(q_C\), \(q_l\) or \(q_{l'}\)}} in this case, the fallback condition will always hold (i.e. \(F(p)=\textrm{\textit{True}}\)). Since it  passes through one of the queues that can feed \(q_{l+1}\) (\(q_C\), \(q_l\) or \(q_{l'}\)), the only effect with respect to the deadlock condition is an earlier falsification of the freedom condition in future packets that actually land in \(q_{l+1}\) (as in case A). This happens when: \vspace{-0.3em} 
\begin{itemize}
\item[--] \(p\)'s destination is the next hop (\(node(q_{l+1})\)). The landing queue will be \(q'_C\), which does not cause a deadlock (being a sink).
\item[--] \(p\) can only go straight for the remainder of its path. The current \(x\) or \(y\)-axis (according to the topology of \(q_l\) and \(q_{l+1}\)) coordinate is the same as the one of the packet destination. This would lead to \(q'_\nearrow\) (\(q'_\uparrow\) for \(F'\)), which already follows the traditional turn model. %
\item[--] 
\(p\) can go to the other forbidden turn \(q'_{l+1}\in\{q'_\lcirclearrowright,  q'_\rcirclearrowleft\}\) or \(\{q'_\lcurvearrowne, q'_\rcurvearrownw,\}\) for \(F'\). 
In this instance, it cannot also have \(q_{l+1}\) as a potential destination (being mutually exclusive to \(q'_{l+1}\)), as U-turns are not allowed in minimal routing. Thus, this remains orthogonal to the studied deadlock condition, though such packets can still go straight, with equivalent outcomes as the instance above. Note that this is subject to the followed turn model, as with \(F'\) that places the forbidden turns on the same downstream router (at North for \(F'\)).
\end{itemize}
\vspace{-0.3em} 
\paragraph*{\hspace{-1em}\textit{C) \(p\) cannot be placed in \(q_C\), \(q_l\) or \(q_{l'}\)}} this case \(p\) includes all other instances when it cannot go towards \(node(q_l+1)\) (at North for \(F'\)). As these are the only queues that feed the queue of the forbidden turn \(q_l+1\), they are unrestricted by following the turn model.\\[-6pt]

In all the aforementioned cases, \(head(q_l)\) would have been the last packet of all queues \(q_l\), \(q_{l'}\) and \(q_{C}\) that could land in \(q_{l+1}\) (for which \(\neg turn(q_{l+1})\)) on the next hop. As \(F(head(q_l))=\textrm{\textit{True}}\) (the now-head) at the time of arrival, it would have been able to be forwarded before \(q_{l+1}\) became full. This packet would have been able to become the tail of \(q_{l+1}\) at least. Thus, when \(q_{l+1}\) is full, the head of \(q_l\) cannot have a dependency to \(q_{l+1}\), i.e. 
\begin{align}
q_{l+1} \notin next(head(q_l)). 
\end{align}
From (3) ( \(q_{l+1} \in next(head(q_l))\) ) and (10), this leads to a contradiction.

\subsection{Multiple-flit packets}\label{mfpr}

When taking into account the existence of multi-flit packets, the cases presented at section \ref{sfpr} can be adapted accordingly. Based on the assumptions, each head-flit contains the packet's length information, which can be used to calculate the worst case occupancy for \(F\) for the entirety of the packets. In other words, the decisions happening at the time a head-flit arrives are enough to ensure the packets fit inside their assigned queues (concerning \(q_{l+1}\)). As this assignment lasts for the entirety of a packet, when \(p\) is not a head flit, it does not consult \(F\) again. It also does not impact the other instances of \(F\) (e.g. from other ports), as its size is considered 0 by the \(occp()\) occupancy function (see section \ref{frco}). 

As an example, packet \(p'\) is another packet (head-flit) that arrives midway through the arrival of a packet/flit \(p\) to a queue from the set \(\{q_C, q_{l'}\}\) (excluding \(q_l\), as output arbiters are required to extract full packets). In this case, all flits from \(p\) are guaranteed to fit inside \(q_{l+1}\), as the freedom condition counts for the whole size of packets. At the same time, \(p'\) is only granted with \textit{F=True} if there is space for it, including all flits of both the packets of \(p'\) and \(p\). %

\section{Evaluation}\label{eval}

The proposed theory is explored by evaluating the proposed novel algorithm examples of section \ref{novela}. First, section \ref{routper} uses high-level simulation to comment on the algorithmic performance of the freedom condition adaptation (\(F'\)). Then, sections \ref{resul} and \ref{resulti} provide implementation-related results based on an example NoC as described with Verilog, a hardware description language (HDL). The goal is to isolate and compare the routing algorithm behaviour as found in the state-of-the-art (see related work in section \ref{rlw}). %

\subsection{Routing performance -- synthetic}\label{routper}

The performance of the proposed routing algorithm methodology is %
studied under a variety of synthetic traffic models in simulation. The presented results are for an \(8 \times 8\) 2D mesh NoC, the routers of which use output queues. The simulation framework builds upon our earlier open source framework for a study on FPGA switches \cite{tpds23switches}.

The traffic models used in this evaluation are selected with diversity in mind, but are also synthetic which is considered the norm for this level of system design at NoC routing. First, uniform Bernoulli arrivals and uniform bursty traffic are the most common models for interconnection circuits and relate to system interconnect use cases \cite{lef, tpds23switches}. Then, bit complement, bit reverse, bit rotate \cite{dally2004principles} and butterfly produce destination permutations based on the corresponding bit manipulation operations on the destination address. These and the transpose model are based on applications, such as the latter for FFT accelerators \cite{dally2004principles}. Finally, hotspot is the Bernoulli model modified for the central node %
to receive requests with four times higher probability than the rest of the nodes, modelling system-on-chip behaviour \cite{dally2004principles,lef}.

\begin{figure*}[h!]
\centering
\includegraphics[width=1\textwidth, trim=0 0 0 0]{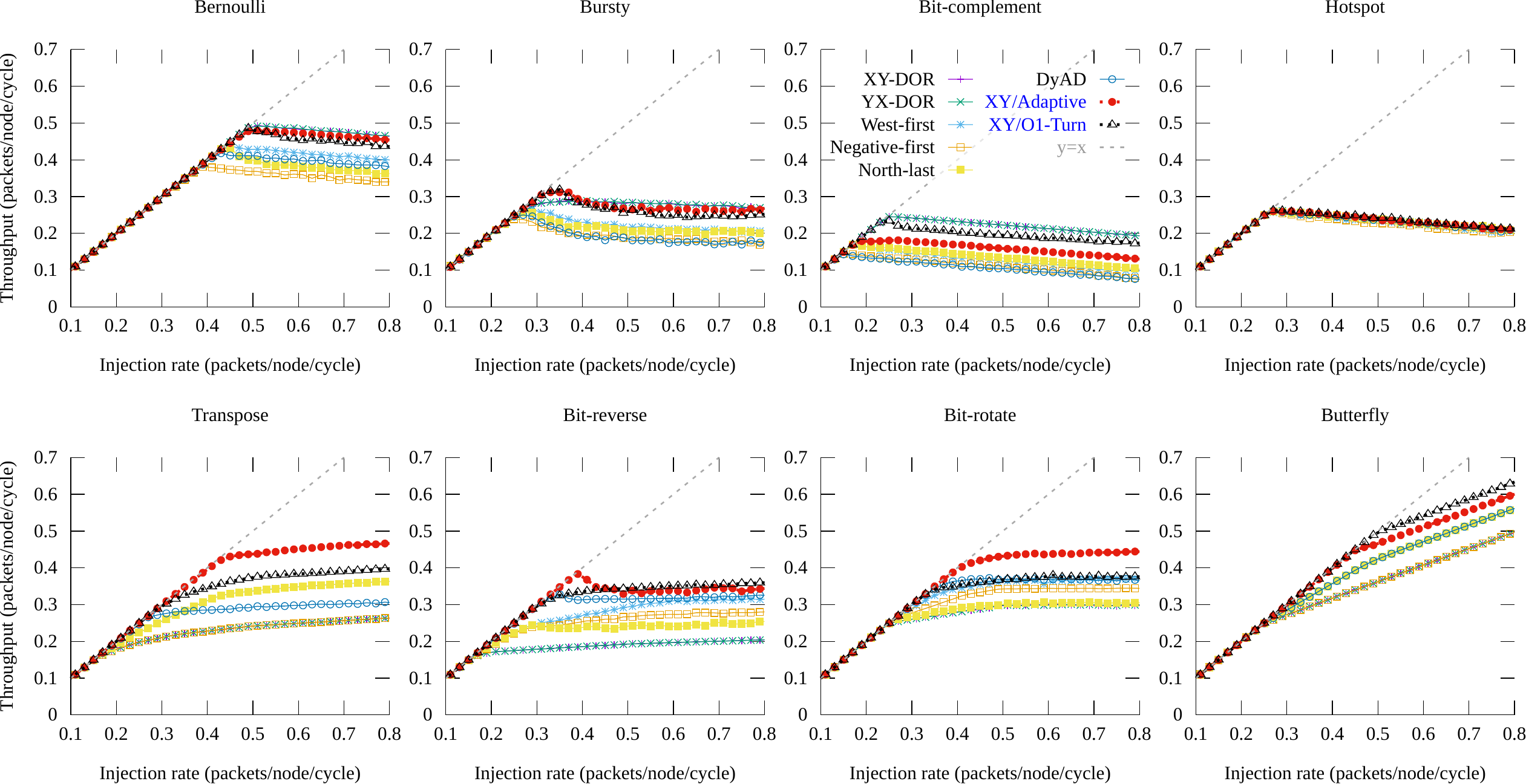}
\caption{Comparison of throughput among different routing algorithms under different traffic patterns}\label{figthr}
\end{figure*}

Figure \ref{figlat} presents the performance results from this experiment with respect to the average packet latency. Each output queue has a depth of 16 flits-packets. There are only single-flit packets and the forwarding from a node can happen under a cycle, also according to the queue states. %
Each simulation has a region of interest of 5,000 cycles. %
warmup period. %
When the average latency is predicted to become above 1,500 cycles, the simulation stops early and the series stops to save simulation time. Each data point is an average of 5 runs. The observations for the equivalent experiment with virtual output queues %
are similar but not shown for brevity.

A general observation is that the novel ``XY/O1-Turn" and ``XY/Adaptive" are the winners in the majority of the traffic patterns, achieving the lowest average packet latency for almost any presented injection rate. Two noticeable exceptions are the uniform Bernoulli and bit-complement cases. In the first, both fully-adaptive algorithms are marginally worse than with dimension order routing (DOR). In the second case, ``XY/Adaptive" comes third, but it is not a close third, so it could be said that `XY/O1-Turn" is a more balanced solution.

The second-class performance of the full-adaptiveness examples under certain traffic cases is expected, but it is not a limitation of the deadlock avoidance model. There are traffic patterns where additional turn-freedom is not always beneficial, at least when the adaptiveness heuristic has a more-local scope \cite{ebrahimi2012catra}. Hence, instead of a fine-tuned routing algorithm, the main focus is the theoretical model that allows a superset of potential packet paths than is currently feasible. For instance, as XY and YX DOR use a subset of the allowable turns, future adaptiveness heuristics could still use the proposed model while also reverting to XY or YX DOR where deemed beneficial to performance. %

Another observation is that overall ``XY/O1-Turn" and ``XY/Adaptive" are more well-rounded than the alternative algorithms in this selection, always being among the top performers. For example, DyAD routing \cite{hu2004dyad} sometimes performs the worst, as under high bursty or bit-complement traffic, whereas for bit reverse and transpose traffic is the next best alternative to the proposed ones.

Figure \ref{figthr} presents the throughput results from the same experiment for numerical comparison examples. %
The throughput is defined as the average number of extracted packets per node per cycle. On average for all the traffic models, under a 35\% injection rate, ``XY/Adaptive" provides 1.23, 1.22, 1.17, 1.28, 1.19 and 1.19x %
the throughput of XY, YX DOR, west-first, negative-first, north-last and DyAD respectively. At 35\% the corresponding numbers for ``XY/O1-Turn" remain very similar at 1.23, 1.22, 1.17, 1.29,  1.19 and 1.17 times.

\subsection{Routing performance -- traces}\label{routper2}

Alongside synthetic workloads, we also conduct a similar experiment with the studied routing algorithms using real-world benchmarks. Specifically, we employ the PARSEC \cite{bienia2011benchmarking} traces provided by Hestness et al. \cite{hestness2010netrace}, which are derived from full-system simulations on the M5 simulator \cite{binkert2006m5}. In our experiments, the packet generation is based on the injection timestamps and traffic patterns (that is, patterns of source-destination pairs) contained in these traces. 

\begin{figure}[h!]
\centering
\includegraphics[width=0.49\textwidth, trim=0 8 0 0]{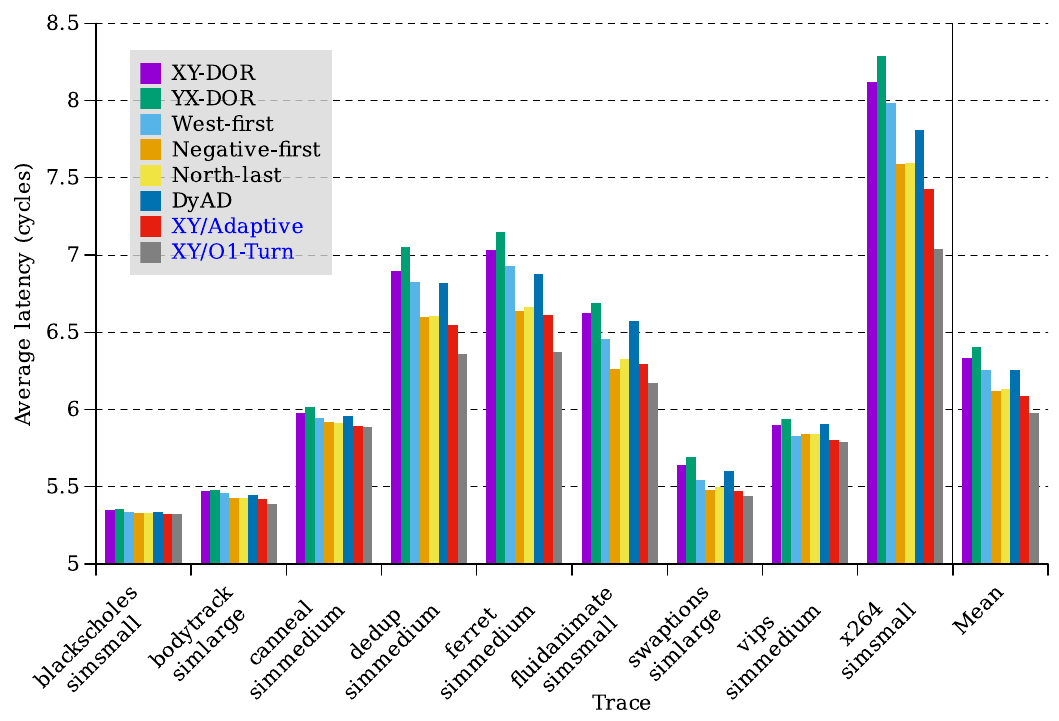}
\caption{Packet latency using different algorithms under traffic from Netrace}\label{trfig}
\end{figure}

Figure \ref{trfig} presents the results of this comparison. Under this traffic, the general observation is that the proposed routing algorithms remain competitive throughout all traces, with ``XY/O1-Turn'' taking the lead with %
up to 8\% lower average latency over the best non-proposed alternative. %

\subsection{Resource utilisation}\label{resul}

In order to study the hardware utilisation of the proposed deadlock avoidance methodology, an \(8\times8\) NoC is implemented in System Verilog and synthesised using yosys \cite{wolf2016yosys}. 

Each NoC router has a \(5\times5\) output-queued switch, with 25 queues in total. The router queues are based on a rather popular formally-verified synchronous queue \cite{bushnyuzi}. The size of each queue is indicatively set to 8, and each packet is 64 bits wide. As building an optimised and specialised implementation is outside the scope of this paper, the router designs do not feature pipelining and the NoC nodes do not perform a useful task (random packet generation). The results are presented for both 6-input lookup tables (LUT6) that are found in most modern FPGAs \cite{lut6}, and a standard cell library (``cmos\_cells.lib'') for ASICs.

Table \ref{tab1} presents the synthesis results for 4 variations of the NoC router based on their routing algorithm. From left to right, each approach is expected to use more resources. North-last uses queue information in its heuristic, while DOR does not. ``Full-freedom" (deadlock-prone) uses a superset of this information, as its heuristic is responsible for the turns from the north as well. This is the equivalent of ``8/8 of turns'' from figure \ref{fimot}. Finally, the proposed example routing algorithm ``XY/Adaptive" includes the signals required for the freedom condition, hence also the two extra wires in the y-axis carrying occupancy information (for queues \(q'_{\rcirclearrowleft}\) and \(q'_\lcirclearrowright\)). This is reflected in the addition of 4 public wires, two as inputs in the north direction and two as outputs of the downstream router for monitoring the two restricted turns.

\begin{table}[h!] 
  \caption{Routing algorithm impact on router resource utilisation} 
\label{tab1} 
\centering
  
 \setlength{\tabcolsep}{3.3pt}
\begin{tabular} {l|r r r r r }

&XY DOR&North-last&\thead{Full-freedom \\ (deadlock-prone)}&\textcolor{darkblue}{XY/Adaptive} \\

\hline
\\[-6pt]
&\multicolumn{4}{c}{\textit{Look-up table-based (FPGA)}}\\
\\[-6pt]
Wires&885&957&977&1002\\
Public wires&182&182&182&186\\
Cells&1132&1204&1224&1245\\
6-LUT&1092&1164&1184&1205\\
\\
&\multicolumn{4}{c}{\textit{Standard cell library-based}}\\
\\[-6pt]
Wires&10707&10985&11230&11458\\
Public wires&183&183&183&187\\
Cells&5630&5797&5962&6062\\
NAND&3476&3803&3566&3823\\
NOR&1455&1344&1649&1518\\
NOT&659&610&707&681\\
\\
&\multicolumn{4}{c}{\textit{Common}}\\
\\[-6pt]
DFF&15&15&15&15\\Sync. FIFOs&25&25&25&25\\

\end{tabular}
\end{table}

It is also worth mentioning that for the %
turn-restriction based (XY DOR and north-last) yosys was not able to optimise unused queues away, as 25 queues were used in all cases. As the proposal is based on the observation that (V)OQs imply turn information, we know that the first two cases could result in (4 for XY DOR, and 2 for north-last) fewer queues. The no-misrouting assumption is also not exploited without manual intervention, as it could yield 5 fewer queues per router (for when sending to oneself). Additionally, as synthesis is also based on heuristics, small variations are expected, such as in the distribution of logic cells in the standard library.

As can be observed, ``XY/Adaptive" being a superset of the ``full-freedom" in terms of logic complexity, it has a small but measurable overhead on the wire and cell utilisation. For the FPGA case, the proposed approach only uses 21 more LUTs %
($<$2\% increase). By using the standard cell library, the most noticeable change is in the NAND gate count that increases by 7\%.

\subsection{Performance overhead}\label{resulti}

This part of the evaluation demonstrates that the performance overhead of including the proposed deadlock avoidance mechanism is minimal. Although the paper mostly provides a theoretical background rather than implementation insights, an indicative NoC implementation verifies our expectations on the impact on the operating frequency.
The last two router implementations from the resource utilisation study (section \ref{resul}) are used as a building block to build a 4x4 NoC on an FPGA. The first of these routers is the deadlock-prone version that enables ``full-freedom'' by having no turn restriction and follows the adaptiveness heuristic to achieve high scheduling performance (see section \ref{novela}). The second one is the ``XY/Adaptive'', which is essentially the same, but with the proposed deadlock avoidance mechanism enabled. The idea is to focus on the impact of adding this additional logic and wires in the NoC's routers, while having all other aspects common, including the adaptiveness heuristic.

The target device is Xilinx UltraScale+ ZU3EG using Avnet's Ultra96 board. The toolchain is Vivado 2020.2 and is used to place-and-route the logic onto the device. The implementation directives are set to  ``ExploreWithRemap'' for design optimisation, and ``Explore'' for post-place and post-route physical design optimisation. The idea is to have relatively aggressive optimisation to minimise the effects of place-and-route heuristic-related discrepancies when reporting the maximal operating frequency (\(f_{max}\)).

The resulting worst negative slack is -1.69 ns for ``full-freedom'', and -1.706 ns for ``XY/Adaptive'', for a target clock of 375 MHz. This results in a sub-1 MHz drop when extending ``full-freedom'''s logic to become ``XY/Adaptive'', as the \(f_{max}\) of the two is about 230 MHz and 229 MHz respectively. 

As mentioned, there can still be small variations based on the nature of place-and-route tools. Still, the numerical results show that the proposed method can be supported efficiently in implemented designs. %
Note that the NoC design space is very narrow, but it also represents a worse case with respect to how the approach can be used. This is since no NoC function is pipelined and all router logic tries to fit inside a single FPGA cycle. See future work (section \ref{ftw}) on a further discussion on implementation.

\section{Related Work}\label{rlw}

Deadlock avoidance, or detection and recovery in NoCs is still an active and thorough research topic. A great majority of the related literature is about NoCs  based on virtual channels (VCs) instead of (virtual) output queues. 

The proposed algorithm is applicable to NoC routers that feature a VOQ-like (VOQ or OQ) queuing topology. On FPGAs, this includes the commonly used split-merge router \cite{huan2012fpga} and variations or adaptations based on the same basic switch primitives \cite{kapre2006packet}. For instance, the Hoplite NoC \cite{kapre2015hoplite} uses the split-merge router, but with less buffering and variations for different NoC topologies \cite{HELAL201851}. All aforementioned examples are classified into the output-queued category, and utilise routing algorithms following the turn model (DOR and west-first). Such switches with static virtual channel allocation by the use of OQ or VOQ queues are focused on implementation efficiency, given certain constraints. Routers based on OQ seem to be the best option for FPGA implementation because there is no need for an FPGA implementation of a crossbar. A crossbar's scheduling algorithm is better represented by hardened circuits, as it is more challenging to include clock domain crossing and iterative algorithms on a unified FPGA logic \cite{tpds23switches}. Hence the natural decision for FPGA-optimised NoCs to use OQ routers.
Our research aims to close %
the knowledge gap on modern deadlock avoidance techniques on these types of NoCs.

In general, deadlock freedom can be achieved by focusing on the routing algorithm or the ﬂow control protocol. A routing algorithm ensures deadlock freedom, if the paths produced by it do not form any cycles. On the other hand, a flow control protocol ensures deadlock freedom if it allocates buffers to packets in a way such that a dependency cycle of packets cannot be formed. Note that this is achieved even in the case the employed routing algorithm is not deadlock-free, that is, it may produce paths that form a cycle.

Glass and Ni conducted a foundational study of deadlock-free routing algorithms for 2D meshes \cite{glass1992turn}. They observed that there are eight possible turns in a 2D mesh and found that cycles of paths can be prevented by prohibiting only two of these eight turns. The details have been described in Section \ref{sec:turn-model}. Chiu \cite{chiu2000odd} pointed out that the flexibility provided by the routing algorithms proposed by Glass and Ni is not even for all source-destination pairs. Specifically, at least a half of the source-destination pairs have only one minimal path while the remaining pairs have more. Chiu proposed to use different sets of prohibited turns for nodes in odd and even columns, called odd-even routing. In this way, the routing flexibility can be improved while deadlock freedom is still guaranteed. 

Our proposed algorithm as illustrated in algorithm \ref{alg2} is inspired by the DyAD routing algorithm \cite{hu2004dyad}. DyAD is a bimodal algorithm as well, as it selects between two component algorithms. One of those components is set as odd-even routing. %
This is provided indicatively, since as with our proposal, it originally models a routing model rather than a fine-tuned algorithm. Nevertheless, the algorithm selection in DyAD only relates to routing performance, and still has permanently forbidden turns. %

There are a number of works also focusing on approaching fully-adaptive routing, but these are still with VCs \cite{ma2013novel,puente2001adaptive,verbeek2011necessary}. Initially, this could be achieved with additional virtual channels called ``escape" channels, where certain packets could resort to in order to avoid deadlocks. Conditional forwarding \cite{yu2016conditional} is a deadlock avoidance mechanism and has similar aspects to our proposed approach. It also uses a function (``conditional forwarding flow control") to determine if a packet can follow a restricted path. It aims to increase flexibility in VC allocation, however. This is achieved by eliminating the need to have separate escape channels, which is a common aspect with our approach on (V)OQs.

\section{Discussion and future work}\label{ftw}
In this section, we discuss the usage aspect of the proposed approach, especially with respect to future adaptations and improvements. The discussion is divided into the themes of \textit{implementation efficiency}, \textit{adaptation} and \textit{improvements}.

\vspace{0.3em} 
\paragraph*{\hspace{-1em}\textit{Implementation efficiency}}There could be additional and/or different simplifications to \(F'\) (see section \ref{fradapt}) 
based on restrictions or architectural assumptions that still satisfy \(F\). One such example could be a quantisation of the queue length measurements for reducing the related signals. %
Another example is to force the 1st hop of all packets to follow the turn model so that \(q_C\) will not need to be checked alongside the other queues that contribute to the contents of \(q'_{\rcirclearrowleft}\) and \(q'_\lcirclearrowright\). The trade-offs between the decisions in such a design space, especially with respect to circuit complexity and algorithm performance would be interesting %
to explore. %

As with other routing algorithms, the model can also be used in pipelined implementations for hardware  scalability and performance. With respect to flow control, a credit system can be used to replace the presented ready signals denoting FIFO space availability \cite{dally2004principles}. 

\begin{figure}[h!]
\centering
\includegraphics[width=0.40\textwidth, trim=0 7 0 7]{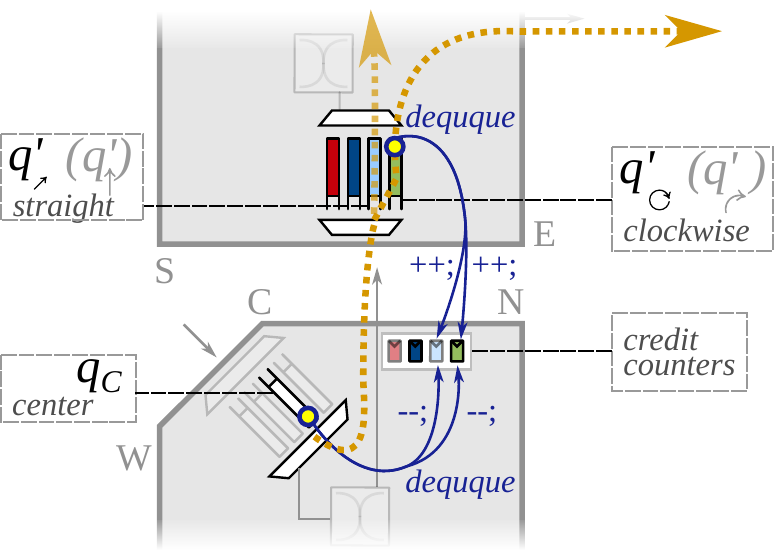}
\caption{Flow control credits can be reused for pipelining \(F\) (shown \(F'\))}\label{ficre}
\end{figure}

Similarly, as shown in figure \ref{ficre}, some of the flow control credit counters can be reused to estimate the freedom condition. The example shows a packet from the upstream router that can pass through \(q'_{\uparrow}\) and the originally-forbidden \(q'_{\lcurvearrowne}\) of the downstream router. The upstream router already maintains a credit counter for each buffer of the downstream router (the buffers in the channel connected to the upstream router) by using a credit-based flow control. When it is dequeued, both of the counters of the corresponding queues shall be greater than 0, and are both decremented by 1. In the downstream router, when the packet is dequeued from either \(q'_{\uparrow}\) or \(q'_{\lcurvearrowne}\), both corresponding counters are incremented by 1. For the purposes of \(F\), the only useful external queue occupancies are for the forbidden turns (\(q'_{\lcurvearrowne}\) for this example), hence only consulting the corresponding credit counters. Equivalent arrangements can be made for the multiple-flit case, but it may still be worth duplicating the credit registers for the two purposes (flow control and \(F\)) according to the pipelining requirements. Earlier notifications are also possible for restoring the non-taken paths, since this information is known at the time of enqueuing. %

\vspace{0.3em} 
\paragraph*{\hspace{-1em}\textit{Adaptation}}The set of system assumptions (section \ref{assum}) %
is partly due to brevity and practicality, such as for FPGA use. %
Some of them can be straightforward to remove. Misrouting can be supported at the expense of additional checks accounting for the additional movements that are now possible. These are the U-turns for when a packet does two consecutive clockwise or counterclockwise turns, as well as when a packet goes back and forth through the same port. This results in additional monitoring in the queue occupancies, which are the supersets of the clockwise and counterclockwise directions plus the backward movement. This results in checking all queues of the upstream router that feed the same port, instead of only including \(q_d\) for either \(d \in \{C, \nearrow, \rcirclearrowleft\}\) or \(d \in \{C, \nearrow, \lcirclearrowright\}\) in \(F(p)\). Similarly, when incorporating interface queues, the formulas can be modified to take the occupancy of additional queues into consideration for calculating the worst case occupancy of the queues of the ``forbidden" turns. %

The approach can also %
be used meaningfully in systems supporting both packets of unknown and known size. This is achieved by assuming \(F(p)=\textrm{\textit{False}}\) on packets \(p\) of unknown size, while still benefiting the others, since the fallback is localised. Such a scheme would be beneficial as the %
majority of NoC packets are single or few-flit  \cite{ma2013novel}, while streaming traffic is commonly split into fixed packets rather than flits, at least according to the protocol. Similarly, as with WPF \cite{ma2012whole}, it is not necessary for the queues to fit all the flits of a packet, but larger packets will tend to follow the fallback routing more easily.

\vspace{0.3em} 
\paragraph*{\hspace{-1em}\textit{Improvement}}The \(F\) condition can be extended to provide additional flexibility. For the queues of the upstream router that feed into originally forbidden turns (e.g. \(\{q_C, q_\nearrow, q_\rcirclearrowleft\}\) for \(q'_\lcirclearrowright\)), the notion of the occupancy could also be restricted to only count for the packets whose destination includes the forbidden turn/queue \(q'_f\) (e.g. \(q'_\lcirclearrowright\)), i.e. 
\(occp (q, q'_f)=\sum_{p \in q \land q'_f \in next(p)}{size(p)}\). %
Future work includes the evaluation of this extension, as the additional checks might lead to implementation complexity tradeoffs. 

Finally, the provided routing algorithms ``XY/Adaptive" and ``XY/O1-Turn" are example uses of the approach. A routing algorithm-focused study could propose fine-tuned variants that combine the best performances that are achieved here per benchmark, for instance.

\section{Conclusions}\label{conc}

This paper relaxes the turn model for building fully-adaptive routing algorithms on NoCs with an output-queued or virtual output-queued router architecture. These two queueing topologies are useful in applications such as in FPGAs, where deadlock avoidance is traditionally achieved using non-fully-adaptive algorithms. The proposed algorithm is a hybrid algorithm, with a fallback component being activated only locally based on the proposed freedom condition. The provided example routing algorithms ``XY/Adaptive" and ``XY/O1-Turn" generally outperform older algorithms significantly in simulation. These presented examples are relatively well-rounded across the traffic model selection, though the proposed approach can be used to build other novel routing algorithms and is easily generalisable. An example implementation of ``XY/Adaptive" is shown to have minimal overhead on resource utilisation and operating frequency over when not featuring deadlock avoidance for full routing freedom.

\section*{Acknowledgement}
\small
The support of the JSPS, Japan KAKENHI Grant Number 21K17720 and the Human Capital Initiative (HCI) of HEA, Ireland is gratefully acknowledged.

\balance
\bibliography{bibliography}

\begin{IEEEbiography}[{\includegraphics[width=1in, height=1.25in,clip, keepaspectratio]{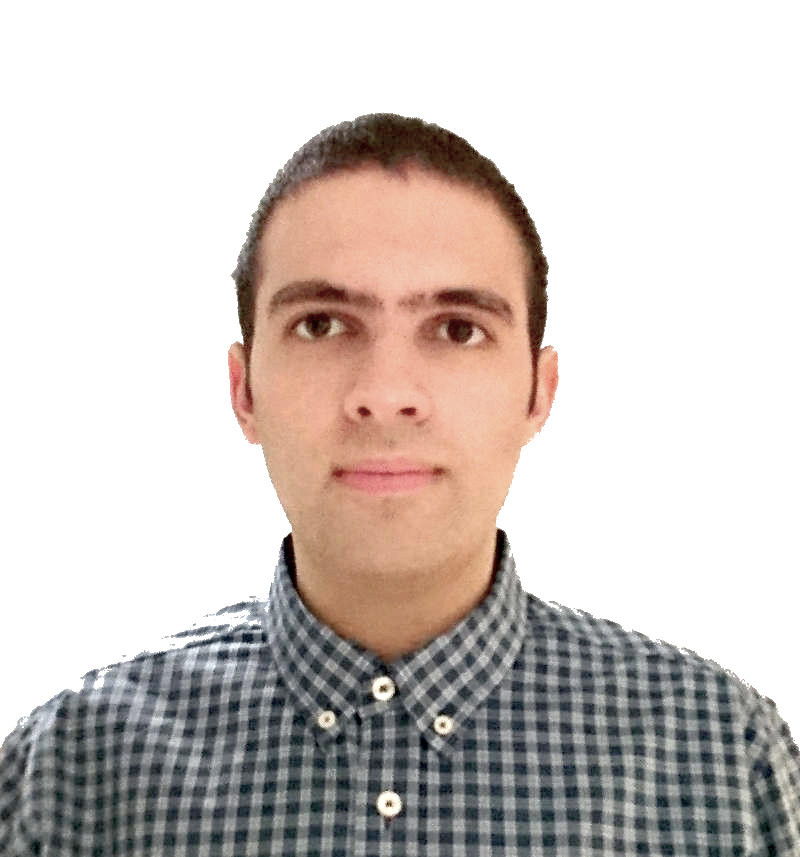}}]{Philippos Papaphilippou}
received his PhD from Imperial College London in 2021. %
His PhD was funded by dunnhumby (Tesco) for researching novel accelerators to improve the
performance of big data analytics. %
He has recently joined Trinity College Dublin as an Assistant Professor for contributing to the Human Capital Initiative (HCI). 
His research topics include FPGAs, sorting algorithms,
network switches, multi-processor architectures and data science.
\end{IEEEbiography}
\begin{IEEEbiography}[{\includegraphics[width=1in, height=1.25in,clip, keepaspectratio]{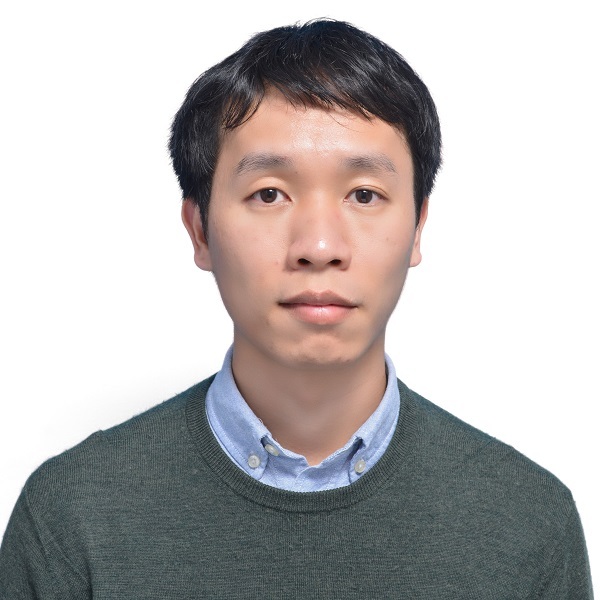}}]{Thiem Van Chu}
completed his Ph.D. at Tokyo Institute of Technology in 2018. Upon graduation, he joined the School of Information Science, Japan Advanced Institute of Science and Technology as an Assistant Professor. In 2020, he moved to Tokyo Institute of Technology. His research interests lie at the intersection of computer architecture, reconfigurable computing, and machine learning.
\end{IEEEbiography}

\end{document}